\documentclass[12pt, draftcls, onecolumn]{IEEEtran}
\usepackage{graphicx,epsfig,amssymb,amsmath}
\usepackage{booktabs}
\newcommand{\sinc}{\mathop{\operation@font sinc}\nolimits}
\usepackage{cite}
\newtheorem{theorem}{Theorem}

\begin{document}

\title{Adaptive Compressive Spectrum Sensing for Wideband Cognitive Radios}

\author{Hongjian~Sun,~\IEEEmembership{Member,~IEEE,} Wei-Yu~Chiu,~\IEEEmembership{Member,~IEEE,}
    and A. Nallanathan,~\IEEEmembership{Senior Member,~IEEE}
\thanks{Copyright (c) 2012 IEEE. Personal use of this material is permitted. However, permission to use this material for any other purposes must be obtained from the IEEE by sending a request to pubs-permissions@ieee.org.}
\thanks{This manuscript has been accepted to be published in IEEE Communications Letters. The associate editor coordinating the review of this letter and approving it for publication was O. Dobre. Digital Object Identifier 10.1109/LCOMM.2012.092812.121648}
\thanks{H. Sun and A. Nallanathan are with the Department of Electronic Engineering, King's College London, London, WC2R 2LS, U.K. (Email:mrhjsun@hotmail.com, nallanathan@ieee.org).}
\thanks{W.-Y. Chiu is with the Department of Electrical Engineering, Princeton University, Princeton, NJ 08544, U.S.A. (Email: chiuweiyu@gmail.com).}
\thanks{The authors acknowledge the support of the UK Engineering and Physical
Sciences Research Council (EPSRC) under Grant No. EP/I000054/1.}}

\vspace{-2em}

\maketitle

\begin{abstract}
This letter presents an adaptive spectrum sensing algorithm that detects wideband spectrum using sub-Nyquist sampling rates. By taking advantage of compressed sensing (CS), the proposed algorithm reconstructs the wideband spectrum from compressed samples. Furthermore, an $\ell_2$ norm validation approach is proposed that enables cognitive radios (CRs) to automatically terminate the signal acquisition once the current spectral recovery is satisfactory, leading to enhanced CR throughput. Numerical results show that the proposed algorithm can not only shorten the spectrum sensing interval, but also improve the throughput of wideband CRs.
\end{abstract}


\section{Introduction}
\label{section1}

Recently, cognitive radio (CR) has attracted much attention due to its capability of exploiting spectral holes and improving spectral utilization efficiency~\cite{ruizhang1, ruizhang2}. This capability is fulfilled by spectrum sensing which is defined as a technique for achieving awareness about the spectral usage and existence of primary users (PUs). With a ``wider'' spectral awareness, CR could exploit more spectral opportunities and achieve greater capacity. Thus, spectrum sensing over wideband spectrum becomes increasingly important for wideband~CRs.

To implement wideband spectrum sensing, CRs need some essential components, i.e., wideband antenna, wideband radio frequency (RF) front-end, and high speed analog-to-digital converter (ADC). The wideband antenna and the wideband filter were well-developed as evidenced by \cite{wantenna} and \cite{wbpf}. By contrast, the development of ADC technology is relatively behind: the achievable sampling rate of the state-of-the-art ADC is only 3.6~Gsps~\cite{ADC}. To deal with this bottleneck, in the classic paper~\cite{scs1}, Tian and Giannakis firstly applied compressed sensing (CS)~\cite{cs} theory to CRs for acquiring wideband signals using sub-Nyquist sampling rates. Consequently, fewer compressed samples are required than predicted on the basis of Nyquist sampling theory. Furthermore, Wang~{\em et al.} \cite{tian3} proposed a two-step CS scheme for minimizing the sampling rate, where the actual sparsity was firstly estimated in the first time slot and the compressed measurements were then adjusted in the second slot. Additionally, Malioutov~{\em et al.}~\cite{scs} studied a sequential CS approach where each compressed measurement was acquired in sequence.

Against this background, the novel contribution of this letter is that an adaptive spectrum sensing algorithm is presented that utilizes CS theory to sense wideband spectrum by using an appropriate number of measurements. Different from the sparsity estimation scheme in \cite{tian3}, the proposed algorithm can adaptively adjust compressed measurements without any sparsity estimation efforts. Instead of the sequential measurement setup in~\cite{scs}, we acquire the wideband signals block-by-block from multiple mini-time slots, and gradually reconstruct the wideband spectrum  using compressed samples until the spectral recovery is satisfactory. The remaining spectrum sensing time slots are utilized for data transmission, thereby enhancing the throughput of wideband CRs. Even with an unknown sparsity level, the proposed algorithm could still automatically terminate signal acquisition at the right time, leading to a robust spectral recovery as well as enhanced CR throughput.

The rest of the letter is organized as follows. Section~\ref{section2} introduces the system model. Section \ref{section3} proposes an adaptive spectrum sensing algorithm. Simulation results are presented in Section \ref{section4}, with conclusions given in Section \ref{section5}.

\section{System Model}
\label{section2}

\begin{figure}[!t]
\centerline{\includegraphics[width=4.4in]{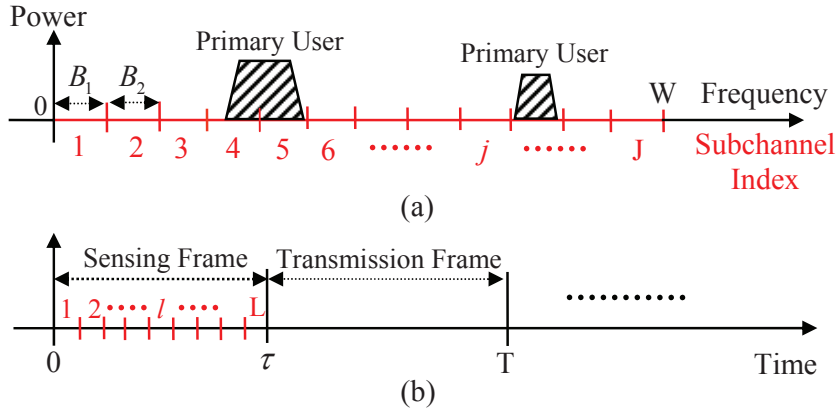}}
\vspace{-0.8em}
\caption{Frequency and time frame in wideband CRs: (a) frequency frame, and (b) time frame. CR employs orthogonal frequency-division multiplexing techniques that divide the wideband spectrum into $J$ subchannels.}
\vspace{-0.4em}
\label{fig0}
\end{figure}

Suppose that CRs aim to exploit spectral holes within frequency band $0\!\sim \!W$ (Hz), as depicted by Fig.~\ref{fig0}(a). Periodic spectrum sensing time frame is adopted as shown in Fig.~\ref{fig0}(b) where $0\sim \tau$ (second) is used for performing spectrum sensing and $\tau \sim T$ (second) is reserved for transmitting data. During the spectrum sensing interval, all CRs keep quiet as enforced by protocols, e.g., at the media access control layer. Thus, the continuous signal received at the RF front-end of CR, i.e., $x_{\textrm{c}}(t)$, is composed of only PUs' signals and background noise. By using sampling rate~$f_{N}$ over the observation time $\tau$, we could obtain a discrete time sequence $x[n]=x_{\textrm{c}}(\frac{n}{f_{N}}),~n=0, 1, \cdots, N-1$, in a vector form $\vec{x}\in \mathbb{C}^{N\times 1}$. Here, $N=\tau f_{N}$ is chosen to be a natural number. After spectrum sensing, CRs adopt orthogonal frequency-division multiplexing (OFDM) techniques that decompose the wideband spectrum into $J$ orthogonal subchannels, each of which has bandwidth $B_j=\frac{W}{J}$ ($\forall~j \in [1, J]$), as shown in Fig.~\ref{fig0}(a). The subchannel index is denoted by $j \in [1, J]$ and PUs may present at any subchannels. For simplicity, let $\Omega$ denote the set of subchannel indices where PUs present. However, based on the Nyquist sampling theory, the sampling rate is required to exceed $2W$ samples per second, i.e., $f_{N}>2W$; for a wideband CR, it leads to excessive memory requirement and prohibitive energy cost. This dilemma motivates us to employ CS technologies to reduce the sampling rate while retaining the spectrum sensing bandwidth $W$.

CS theory indicates that, if a signal is sparse in some basis, it can be acquired by using a sub-Nyquist sampling rate; thus, fewer compressed samples are obtained than predicted using the Nyquist sampling theory. Mathematically, by using sub-Nyquist sampling rate~$f_{S}$ ($f_{S}<2W$), the compressed samples $\vec{y}$ ($\vec{y}\in \mathbb{C}^{M\times 1}$, $M=\tau f_{S} \ll N$) can be written as
\begin{equation}
\vec{y}=\mathbf{\Phi} \vec{x}
\end{equation}
where $\mathbf{\Phi}$ denotes an $M\times N$ measurement matrix. Notably, Tropp {\em et al.}~\cite{beyond} cleverly implemented a CS system in which the measurement matrix is known and adjustable by changing pseudo-random sequences. For a comprehensive understanding of CS implementation, the reader is referred to~\cite{beyond}.

In a CS-based spectrum sensing system, the goal is to reconstruct $\vec{x}$ or its discrete Fourier transform (DFT) spectrum $\vec{X}=\mathbf{F}\vec{x}$ ($\mathbf{F}$ denotes a DFT matrix) from $\vec{y}$. Then the traditional spectrum sensing algorithm, e.g., energy detection~\cite{hongjian}, can be used to perform spectrum sensing using the reconstructed signal. For a robust signal recovery, the vector $\vec{x}$ is required to be sparse in some basis. Due to low spectral occupancy, it is believed that the received signal at CRs is sparse in the Fourier domain~\cite{scs1}. Thus, $\vec{x}$ is often assumed to $k$-sparse ($k<M\ll N$) in the Fourier domain, which means that the DFT spectrum $\vec{X}$ consists of $k$ significant components which are not negligible. If this spectral sparsity level $k$ is known, we can choose the number of measurements $M$ to secure the quality of spectral recovery, e.g., $M=C_0 k\log (N/k)$ for a Gaussian measurement matrix, where $C_0$ denotes a constant~\cite{cs}. Nevertheless, in a practical CR system, the spectral sparsity level is often unknown or difficult to estimate due to the dynamic activities of PUs. Furthermore, to avoid incorrect spectral recovery, traditional CS approaches tend to pessimistically choose $C_0$. Both phenomena can lead to more number of measurements and higher energy consumption, therefore, losing the advantage of using CS technologies.

\section{Adaptive Spectrum Sensing}
\label{section3}

In this section, we study an adaptive spectrum sensing algorithm for wideband CRs.

\subsection{System Description}
\label{section3.1}

Consider that a CS system, e.g., random demodulator~\cite{beyond}, is employed for implementing wideband signal acquisition as the discussions in Section~\ref{section2}. Rather than sampling the wideband signal for the whole spectrum sensing interval in the traditional CS system, we propose to acquire the wideband signal step by step. The proposed algorithm aims to terminate the signal acquisition once the spectral recovery is satisfactory, and use the remaining spectrum sensing time interval for data transmission. The detailed algorithm is given in Table~\ref{table:a1}.

\begin{table}[!t]
\centering
\tabcolsep 3pt
\caption{Adaptive Spectrum Sensing Algorithm.} 
\begin{tabular}{l}
\hline 
\hline
\toprule
{\bf Initialize:} Divide the spectrum sensing interval $\tau$ into $L$ mini time slots \\
\hspace{4em} and set the mini time slot index $l=1$ and accuracy $\varepsilon$.\\[0.5ex]
\hline
\midrule
{\bf While} the halting criterion is false and $l\le L$, {\bf do}\\[0.5ex]
\hspace{2em} a). Acquire the compressed samples till the mini time slot $l$, \\
\hspace{3.5em} resulting in the set of compressed samples $\vec{y}_{l}$.\\
\hspace{2em} b). Decompose the compressed samples $\vec{y}_{l}$ into \\
\hspace{3.5em} the training subset $\vec{R}_{l}$ and the testing subset $\vec{V}_{l}$.\\
\hspace{2em} c). Estimate the wideband spectrum by applying a certain recovery \\
\hspace{3.5em}  algorithm to (\ref{solve}), leading to a spectral estimate $\hat{X}_{l}$.\\
\hspace{2em} d). Calculate $\rho_l$ by using $\vec{V}_{l}$ and (\ref{very}). \\
\hspace{2em} e). {\bf If} the halting criterion is true \\
\hspace{4em} 1). Terminate the signal acquisition. \\
\hspace{4em} 2). Perform spectrum sensing using $\hat{X}_l$. \\
\hspace{4em} 3). Choose frequency bands and start data transmission. \\
\hspace{3.5em} {\bf Else}:~ $l=l+1$.\\
\hspace{3.5em} {\bf EndIf}\\[0.5ex]
\hline
\midrule
{\bf Halting criterion:} $\left|\rho_{l}/v_{l}- 2\delta^2\right| \le \varepsilon$.\\
\bottomrule
\hline
\hline
\end{tabular}
\vspace{-0.5em}
\label{table:a1} 
\end{table}

As shown in Fig.~\ref{fig0}(b), the spectrum sensing interval is divided into $L$ mini time slots where $l$ ($l \in [1, L]$) denotes the mini time slot index. Let $\vec{y}_{l}$ ($\vec{y}_{l} \in \mathbb{C}^{M_{l}\times 1}$) denote the set of compressed samples obtained from the beginning of spectrum sensing to the end of $l$-th mini time slot, and $M_{l}$ denote the total number of measurements in $\vec{y}_{l}$, thus, $0<M_{1}< \cdots < M_{L}$. Additionally, the sub-Nyquist sampling rate $f_S$ is chosen such that $M_{L}=f_S \tau=C_0 k_{\max}\log (N/k_{\max})$ where $k_{\max}$ denotes the maximum sparsity that can be estimated by long-term spectral observations.
The set of compressed samples $\vec{y}_{l}$ is then divided into two complementary subsets, i.e., the training subset $\vec{R}_{l}$ ($\vec{R}_{l} \in \mathbb{C}^{r_{l}\times 1}$) for reconstructing the DFT spectrum, and the testing subset $\vec{V}_{l}$ ($\vec{V}_{l} \in \mathbb{C}^{v_{l}\times 1}$) for validating the spectral recovery, where $M_{l}=r_{l}+v_{l}$.
According to CS theory, the training subset and the testing subset can be written as
\begin{equation}
\vec{R}_{l} =\mathbf{\Phi}_{l} \vec{x}_{l}+\vec{n}=\mathbf{\Phi}_{l} \mathbf{F}^{-1} \vec{X}_{l}+\vec{n}
\label{train}
\end{equation}
and
\begin{equation}
\vec{V}_{l} =\mathbf{\Psi}_{l} \vec{x}_{l}+\vec{n}=\mathbf{\Psi}_{l} \mathbf{F}^{-1} \vec{X}_{l}+\vec{n}
\label{test}
\end{equation}
respectively, where $\mathbf{F}^{-1}$ is the inverse of DFT matrix, $\mathbf{\Phi}_{l}$ is a $r_{l}\times N$ measurement matrix, $\mathbf{\Psi}_{l}$ is a $v_{l}\times N$ testing matrix, and $\vec{n}$ denotes the measurement noise modeled by circular complex additive white Gaussian noise (AWGN) with zero mean and variance $\delta^2$, i.e., $\vec{n} \sim \mathcal{CN}(0, \delta^2)$.

By using a recovery algorithm, e.g., Log-barrier approach~\cite{Log}, we could obtain a spectral estimate $\hat{X}_{l}$ by solving the following problem:
\begin{equation}
\min \| \hat{X}_{l} \|_1,~\textrm{s.t.:}~\| \vec{R}_{l}-\mathbf{\Phi}_{l} \mathbf{F}^{-1} \hat{X}_{l}\|_2 \le \epsilon
\label{solve}
\end{equation}
where $\epsilon$ is a small recovery error threshold. Repeating this procedure, a sequence of spectral estimates, i.e., $\hat{X}_{1}, \hat{X}_{2}, \cdots, \hat{X}_{l}$, will be obtained by increasing the total number of measurements $M_{l}$. Obviously, we would like to identify a ``best'' spectral estimate $\hat{X}_{l}$ that makes the spectral recovery error $\|\vec{X}_{l}-\hat{X}_{l} \|_2$ sufficiently small. If so, we can terminate the signal acquisition, and improve the throughput of CR system by using the remaining spectrum sensing time slots for transmitting data. However, the spectral recovery error $\|\vec{X}_{l}-\hat{X}_{l} \|_2$ is typically unknown due to the unknown $\vec{X}_{l}$ when performing sub-Nyquist sampling. Hence, for a traditional CS system, the signal acquisition cannot be terminated at the right time.

To identify the best spectral estimate, we propose to use the testing subset $\vec{V}_{l}$ for verifying the spectral estimate~$\hat{X}_{l}$. Specifically, we define the following verification parameter:
\begin{equation}
 \rho_{l}=\| \vec{V}_{l}- \mathbf{\Psi}_{l} \mathbf{F}^{-1} \hat{X}_{l}\|_2^2.
 \label{very}
\end{equation}
As we will see in the next section, if the verification parameter $\rho_{l}$ is close enough to $2\delta^2 v_{l}$, the spectral estimate~$\hat{X}_{l}$ is the best spectral estimate and the signal acquisition can be terminated.

\subsection{Performance Analysis and Comparison}
\label{section3.2}

The termination metric in the preceding section is due to the fact that the best spectral estimate can be identified by validating the spectral estimate sequence and monitoring $\rho_{l}$.

\begin{theorem}
\label{theorem1}
 Let $\varepsilon>0$, $\varrho \in (0,1)$, and $v_{l}>0$. Given the sequence of spectral estimates $\hat{X}_1, \cdots, \hat{X}_{l}$, the best spectral estimate exists and is included in the sequence when the verification parameter $\rho_l$ satisfies
\begin{equation}
\Pr \left[ \left|\rho_{l}/v_{l}- 2\delta^2\right| \le \varepsilon \right] > 1-\varrho \label{t2}
\end{equation}
where $\varrho=2\exp \left(-\frac{3 v_{l} \varepsilon^2}{24 \delta^4 + 2(U^2+\delta^2)\varepsilon} \right)$, in which $U$ denotes the measurement noise upper bound, i.e., $\vec{n} \le U$.
\end{theorem}

The proof of Theorem~1 is given in the Appendix.

\emph{Remark 1:} It shows that, if the best spectral estimate exists within a given sequence of spectral estimates, the verification parameter should be within a certain small range around $2\delta^2$ with a high probability. This probability exponentially increases as the size of testing subset increases. In other words, if we monitor~$\rho_l$, we have a higher probability of identifying the best spectral estimate when using more measurements for validation. However, in another application scenario where the total number of measurements $M_l$ is fixed, there exists a trade-off between training and testing. Even though allocating more measurements for validation (i.e., $v_{l}$) achieves a higher probability of identifying the best spectral estimate,  it leads to a degraded probability of successful spectral recovery because of fewer measurements for training (i.e., a larger $v_{l}$ leads to a smaller $r_{l}=M_l-v_{l}$). The investigation of this trade-off is an interesting issue for future research.

Suppose that the signal acquisition is terminated at mini time slot~$l^{\star}$, then the remaining time slots $l^{\star}+1, \cdots, L$ could be used for transmitting data. Thus, the aggregate opportunistic throughput of the proposed CR system can be given by
\begin{equation}
C^{\star}=\frac{T-\frac{\tau }{L}l^{\star}}{T} \sum_{j\notin \Omega} (1-P_{\textrm{f},j})\cdot B_j \cdot \log \left( 1+ \frac{P_j |H_j|^2}{N_0 B_j}\right)
\label{prop}
\end{equation}
where $P_{\textrm{f},j}$ is the probability of false alarm, $P_j$ is the transmit power of CR transmitter, $H_j$ denotes the magnitude channel gain between the CR transmitter and the CR receiver at subchannel $j$, and $N_0$ denotes the noise spectral density. By contrast, a traditional CS system, the aggregate opportunistic throughput of CR system is given by
\begin{equation}
C=\frac{T-\tau}{T} \sum_{j\notin \Omega} (1-P_{\textrm{f},j}) \cdot B_j \cdot \log \left( 1+ \frac{P_j |H_j|^2}{N_0 B_j}\right).
\end{equation}
It can be easily seen that the proposed system has superior performance than the traditional system due to $l^{\star} \le L$ in (\ref{prop}).

\section{Numerical Results}
\label{section4}

In simulations, we consider the following wideband signal:
\begin{eqnarray}
x_{\textrm{c}}(t)\!=\!\mathop{\sum}\limits_{j=1}^{N_b}\! \sqrt{E_j B_j} \textrm{sinc} \! \left(B_j(t \! -\! \alpha)\right) \! \cos \left(2\pi f_{j} (t\! -\! \alpha) \right)\!+\!z(t),
\end{eqnarray}
where sinc$(x)=\frac{\sin (\pi x)}{\pi x}$, $\alpha$ is a random time offset, $z(t)$ is AWGN with zero mean and unit variance, and $E_j$ is the received power at CR at subband $j$. The wideband signal consists of $N_b=8$ non-overlapping subbands. The subband $j$ is in the frequency range [$f_{j}-\frac{B_j}{2}$, $f_{j}+\frac{B_j}{2}$], where the bandwidth $B_j=10\sim 30$~MHz and the center frequency $f_{j}$ is randomly located in $[\frac{B_j}{2}, W-\frac{B_j}{2}]$ in which the overall bandwidth $W=2$ GHz. The received signal-to-noise ratios (SNRs) of these 8 active subbands are random natural numbers between 7~dB and 27~dB. One time frame has length $T=10$~$\mu$s, in which the spectrum sensing interval is $\tau=5$~$\mu$s. The spectrum sensing interval is divided into $L=20$ mini time slots.
Rather than using the Nyquist sampling rate $f_{N}=2W=4$~GHz, we adopt the sub-Nyquist sampling rate $f_{S}=1$~GHz. The number of compressed samples in a traditional CS system is $M=f_{S}\tau=5,000$, whereas $N=f_{N}\tau=20,000$. The measurement matrix and the testing matrix follow the standard normal distribution with zero mean and unit variance. The measurement noise is assumed to be circular complex AWGN, i.e., $\vec{n} \sim \mathcal{CN}(0, \delta^2)$. The signal-to-measurement-noise ratio (SMNR) is set to~$50$ dB.  The energy detection approach in \cite{hongjian} is employed to detect PUs by using the reconstructed spectrum. For the data transmission, CRs adopt the transmit power $P_j=30 \sim 50$ dBm. The channel between the CR transmitter and the CR receiver is assumed to be block slow fading channel with the path loss given by the 3GPP simulation guideline~\cite{3gpp}: $127+30 \log_{10} (D)$, where $D$ (km) denotes the distance between the CR transmitter and the CR receiver.

\begin{figure}[!t]
\centerline{\includegraphics[width=4.4in]{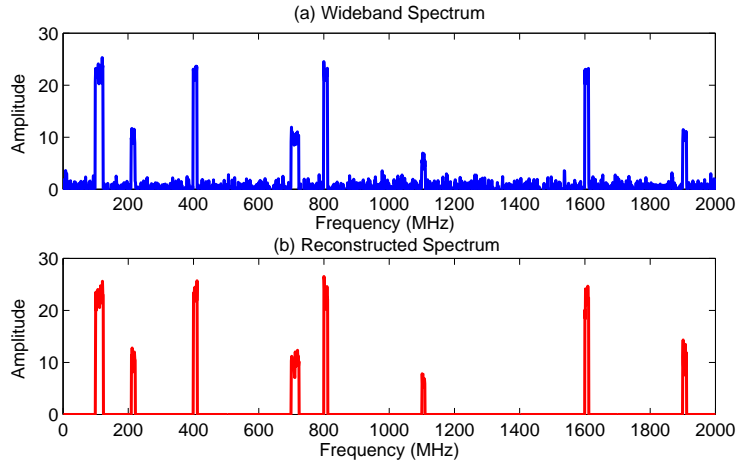}}
\vspace{-0.8em}
\caption{Examples of: (a) the wideband spectrum $\vec{X}$, and (b) the reconstructed spectrum $\hat{X}$. The signal acquisition is terminated at mini time slot 10, with the number of compressed samples $2,500$. The SNRs of these 8 active subbands were set to random natural numbers between 7~dB and 27~dB. }
\vspace{-0.4em}
\label{fig3}
\end{figure}

\begin{figure}[!t]
\centerline{\includegraphics[width=4.4in]{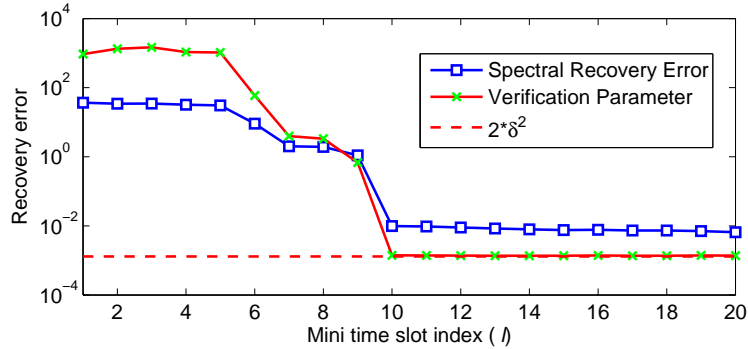}}
\vspace{-0.8em}
\caption{Comparison of the verification parameter $\rho_{l}/v_{l}$ and the predicted value $2 \delta^2$. The actual spectral recovery error is also shown.}
\vspace{-0.4em}
\label{fig2}
\end{figure}

As we can see from Fig.~\ref{fig3}, the wideband signal is composed of both high SNR subbands and low SNR subbands. Using the proposed algorithm, we can successfully reconstruct the wideband spectrum and terminate the signal acquisition at mini time slot~10, instead of mini time slot $L=20$ when using the traditional CS algorithm. This is because, as shown in Fig.~\ref{fig2}, the verification parameter becomes very close to $2\delta^2$ just when the (unknown) actual spectral recovery error becomes sufficiently small. Hence, if the result of Theorem~1 is used as the signal acquisition termination metric, the issue of excessive numbers of measurements can be solved. Fig.~\ref{fig4} shows that the wideband CR using the proposed algorithm outperforms the CR system using the traditional CS algorithms. The throughput gain improves as the transmit power increases. The reason is that, even with the same sub-Nyquist sampling rate, the proposed algorithm utilizes less time slots for performing spectrum sensing than that of traditional CS algorithms.

\section{Conclusions}
\label{section5}

In this letter, we have proposed an adaptive spectrum sensing algorithm for improving the throughput of wideband CRs using CS technologies. It has been shown that the proposed algorithm can successfully reconstruct the wideband spectrum by using a few sub-Nyquist samples. Additionally, the wideband signal acquisition can be automatically terminated even if the actual spectral recovery error is unknown, thanks to the $\ell_2$ norm validation approach.
Furthermore, it has been proved that the proposed CR system can provide greater throughput than the CR system using traditional CS technologies. Simulation results have shown that the proposed algorithm can not only adaptively reconstruct the wideband spectrum by using an appropriate number of measurements, but also offer enhanced throughput for wideband CRs.


\appendix[Proof of Theorem~1]
Due to the parameter setting in Section~\ref{section3.1}, the best spectral estimate (with zero or sufficiently small $ \| \vec{X}_{l} -\hat{X}_{l}\|_2$) should exist and be included in the sequence $\hat{X}_{1}, \hat{X}_{2}, \cdots, \hat{X}_{L}$. Assuming $\hat{X}_{l} (l \in [1, L])$  is the best spectral estimate, the verification parameter $\rho_{l}$ can be written as
\begin{eqnarray}
 \rho_{l} \!\!&\!\!=\!\!&\!\!\| \vec{V}_{l}- \mathbf{\Psi}_{l} \mathbf{F}^{-1} \hat{X}_{l}\|_2^2= \| \mathbf{\Psi}_{l} \mathbf{F}^{-1} (\vec{X}_{l}-\hat{X}_{l})+\vec{n} \|_2^2 \nonumber \\
 \!\!&\!\!\approx\!\!&\!\! \|\vec{n}\|_2^2= \sum_{i=1}^{v_{l}} (n_{R,i}^2+n_{I,i}^2)
\label{valid1}
\end{eqnarray}
where $n_{R,i}$ and $n_{I,i}$ denote the real and imaginary parts of the measurement noise, respectively. As $n_{R,i}$ and $n_{I,i}$ are normally distributed with zero mean and variance $\delta^2$, we obtain $\mathbb{E}(n_{R,i}^2)=\mathbb{E}(n_{I,i}^2)=\delta^2$, and $\textrm{Var}(n_{R,i}^2)=\mathbb{E}(n_{R,i}^2\!-\!\delta^2)^2=\textrm{Var}(n_{I,i}^2)=\mathbb{E}(n_{I,i}^2\!-\!\delta^2)^2=2\delta^4$. Additionally, we find $|n_{R,i}^2-\delta^2|\le |n_{R,i}|^2+\delta^2 \le U^2+\delta^2$.

Applying the Bernstein's inequality~\cite{bein}, we can obtain
\begin{eqnarray}
\!\! \!\!\!\!&\!\!\!\! &\!\!\!\!\!\!  \Pr \!\left[ \!\left| \!\sum_{i=1}^{v_{l}} \!\!(n_{R,i}^2\! + \!n_{I,i}^2) \!-\! 2\delta^2 v_{l} \! \right| \!> \!\xi \!\right] \!\!=\!\!\Pr \! \left[\! \left|\!\sum_{i=1}^{v_{l}} \!\!(n_{R,i}^2\!-\!\delta^2\! + \!n_{I,i}^2\!-\!\delta^2 \!) \!\right| \!>\! \xi \!\right] \nonumber \\
\!& \!\!\!\le \!\!\!&\!\!\!2\exp \! \left(\!-\frac{\xi^2/2}{\sum \mathbb{E}(n_{R,i}^2\!-\!\delta^2)^2 \!+ \!\sum \mathbb{E}(n_{I,i}^2\!-\!\delta^2)^2 \!+ \! (U^2\!+\!\delta^2)\xi/3} \right)\nonumber \\
 \!& \!\!\!\le \!\!\!& \!\!\! 2\exp \left(-\frac{3 \xi^2}{24 \delta^4 v_{l} + 2(U^2+\delta^2)\xi} \right).
 \label{inq}
\end{eqnarray}
Considering both (\ref{valid1}) and (\ref{inq}), we obtain
\begin{eqnarray}
\Pr \left[ \left|\rho_{l}- 2\delta^2v_{l}\right| \le \xi \right] > 1\!-\!2\exp \!\left(\!-\frac{3 \xi^2}{24 \delta^4 v_{l} + 2(U^2+\delta^2)\xi} \right). \label{proof}
\end{eqnarray}
Replacing $\xi$ by $\varepsilon v_{l}$ in (\ref{proof}), we complete the proof.

\begin{figure}[!t]
\centerline{\includegraphics[width=4.4in]{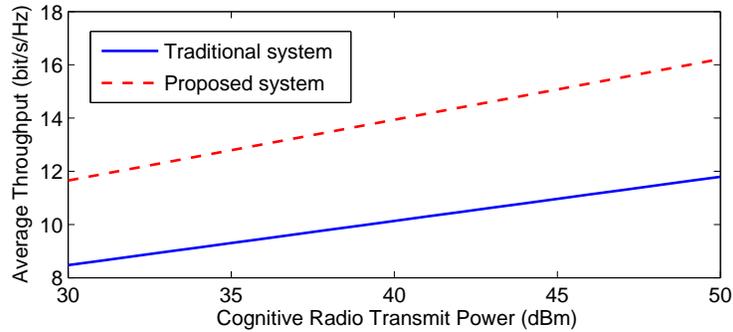}}
\vspace{-0.8em}
\caption{Performance comparison of the proposed wideband CR system and the wideband CR system based on traditional CS when the distance $D=50$~m.}
\vspace{-0.4em}
\label{fig4}
\end{figure}

\bibliographystyle{IEEEtran}
\bibliography{cwpf}

\end{document}